\documentclass[12pt,preprint]{aastex}

\newcommand\kms{~\rm{km\ s^{-1}}}
\newcommand\sigHI{\sigma_{\rm HI}}
\newcommand\HI{H{\footnotesize~I\ }}

\newcommand\HII{H{\footnotesize~II\ }}
\newcommand\GSH{GSH~138$-$01$-$94~}

\slugcomment{To appear ApJ 000,000.}

\shorttitle{GSH 138-01-94}
\shortauthors{J.M. Stil et al.}

\begin{document}

\title{\GSH, \\
    an old supernova remnant in the far outer Galaxy}

\author{J.M. Stil \and J.A. Irwin}
\affil{Queen's University, Kingston, ON K7L 3N6, Canada}
\email{stil@astro.queensu.ca}\email{irwin@astro.queensu.ca}

\begin{abstract}
The properties of the Galactic \HI shell \GSH are derived from data of
the Canadian Galactic Plane Survey.  The basic parameters of \GSH were
determined by fitting the expansion of a thin shell to the expansion
velocity field on the sky. The kinematic distance is $16.6\ \rm kpc$
for $v_{\rm LSR}=-94.2 \pm 0.5 \kms$. The radius is $180 \pm 10\ \rm
pc$, the expansion velocity $v_{\rm exp} = 11.8 \pm 0.9 \kms$, and the
mass $2 \times 10^5\ \rm M_\sun$. No radio continuum counterpart of
the shell was detected at 21 cm or at 74 cm. Absorption of a
background continuum source constrains the spin temperature of \HI in
the shell to $T_{\rm s}=230^{367}_{173}\ \rm K$. The expansion age of \GSH
is $4.3\ \rm Myr$. These observables are in excellent agreement with
predictions from hydrodynamic models for a supernova remnant in a
low-density low-metallicity environment such as the outer Galaxy. \GSH
is then the largest and the oldest supernova remnant known. It
provides direct evidence for the release of mechanical energy in the
interstellar medium by stars in the outer galaxy.  It is argued that
such old supernova remnants be found in low-density, low-metallicity
environments such as the outer Galaxy, dwarf galaxies and low surface
brightness galaxies.
\end{abstract}


\section{Introduction}

The rate and the location of star formation is an important factor in
the energy budget of the interstellar medium and, in the long term,
the evolution of galaxies. The gaseous disk in most galaxies is much
more extended than the stellar disk, leaving a significant fraction of
the mass of the interstellar medium at large radii outside the bright
optical disk. Despite the significant gaseous mass at large radii,
little or no star formation occurs outside the stellar disk.

The most widely accessible tracer of kinetic energy in the
interstellar medium inside and outside the stellar disk of galaxies is
the \HI velocity dispersion, $\sigHI$. Remarkably, $\sigHI$ remains
constant, approximately $10 \kms$, with distance from the centre,
despite a drop of more than two orders of magnitude in the surface
density of stars \citep{vanzee99}.  Energy sources that have been
proposed to sustain the $\sim 10 \kms$ \HI velocity dispersion in the
outskirts of galaxies include previously undetected low-level star
formation \citep{ferguson98a}, magneto-hydrodynamic instabilities in
the disk \citep{sellwood99}, and infalling gas clouds
\citep{hunter98,hunter99}. Which of these, if any, is the dominant
source of energy is still an open question.

In this paper we present evidence for supernova activity in the
extreme outer region of our Galaxy.  Evidence for star formation at
large distances from the Galactic centre exists from detection of
molecular gas, distant O and B stars, and their associated \HII
regions. In this paper, we present evidence for an expanding \HI shell
in the same area of the Galaxy as surveyed by \citet{digel94}. The
structure is visible in the Maryland-Greenbank survey map shown by
\citet{digel94}, but the southern half is so faint that it was not
previously recognized as a shell.

\section{Observations}

\begin{figure}
\caption{ \HI channel maps from the DRAO 21-cm line data. Every other 
channel between $-78\kms$ and $-110\kms$ is shown.
\label{chanmap-fig}
}
\end{figure}

\begin{figure}
\caption{ Latitude-$v_{\rm LSR}$ diagram, constructed by adding three 
latitude-velocity slices at longitude $l=137\fdg865$ and $3\arcmin$
offset in longitude on either side. At this longitude, which is slightly 
offset with respect to the centre of the shell, the slices intersect 
gas at the most extreme velocity. The ellipse represents the thin shell
model fitted in section~\ref{properties-sec}, taking into account that 
the shell is not intersected through the centre in this Figure. 
\label{shellxv-fig}
}
\end{figure}

\subsection{The Canadian Galactic Plane Survey}

The data presented in this paper were obtained as part of the Canadian
Galactic Plane Survey. \HI 21-cm line observations were carried out
with the Dominion Radio Astrophysical Observatory (DRAO)
interferometer, supplemented with zero-spacing information from the
DRAO 26m telescope.  These data represent all structure down to the
$1\arcmin \times 1\arcmin/\cos \delta$ resolution limit of the DRAO
interferometer at 21 cm wavelength and declination
$\delta$. Additional observations of 21-cm continuum and 74-cm
continuum from DRAO, and CO spectral line data from the Five College
Radio Astronomy Observatory (FCRAO) were used.  A description of these
data sets can be found elsewhere
\citep{higgs99,taylor99,heyer98,carpenter97}.

\section{Results}
\subsection{General morphology}
\label{properties-sec}

Figure~\ref{chanmap-fig} shows the shell that we refer to as \GSH, as
a circular structure with radius $37\farcm3$, best visible in the
channel maps around velocity $-96 \kms$.  The blue-shifted side of the
shell represents the most extreme velocity of all \HI in this
direction. On the redshifted side of the shell, the confusion with the
background is stronger.  The channel maps between $-88.0 \kms$ and
$-102.9 \kms$ display a distinct empty region surrounded by areas of
higher intensity. The redshifted half of the shell is clearly fainter
than the blueshifted half. However, the redshifted half shows several
filamentary structures similar to blueshifted filaments that can be
safely attributed to the shell. Examples of such features are found at
$(l,b,v)=(138\fdg3,-1\fdg8,-89.7)$ and ($138\fdg0$,$-0\fdg7$,$-88.0$).
At an even more redshifted velocity the empty region inside the ring
fills up to become a local maximum, most clearly visible in the
channel maps of $-81.4 \kms$ and $-83.1 \kms$, that disappears as the
velocity decreases to $-78 \kms$. On the blueshifted side, the ring
shape starts filling up in the channel maps at $-106.2
\kms$. Therefore, the change in morphology with velocity is symmetric
with respect to the channel map at $-94.6 \kms$, although the
redshifted side is somewhat fainter than the blueshifted side. This
behavior is characteristic of an expanding shell, with both the near
(blueshifted) and far (redshifted) sides visible. At the lowest
intensity levels visible in Figure~\ref{chanmap-fig}, the shell is
nearly continuous, but the side that faces the Galactic equator is a
factor 2 brighter than the opposite side, on average.

The shell is further illustrated in the latitude-velocity diagram in
Figure~\ref{shellxv-fig}. The redshifted emission that we associate
with \GSH traces the doppler ellipse very closely. Furthermore, this
emission, with velocity $-83 \kms$ is restricted to the latitude $b
\approx -1\fdg25$. The emission is therefore not associated with a
more widely distributed component, like most of the emission in
Figure~\ref{shellxv-fig}.

\begin{figure}
\plottwo{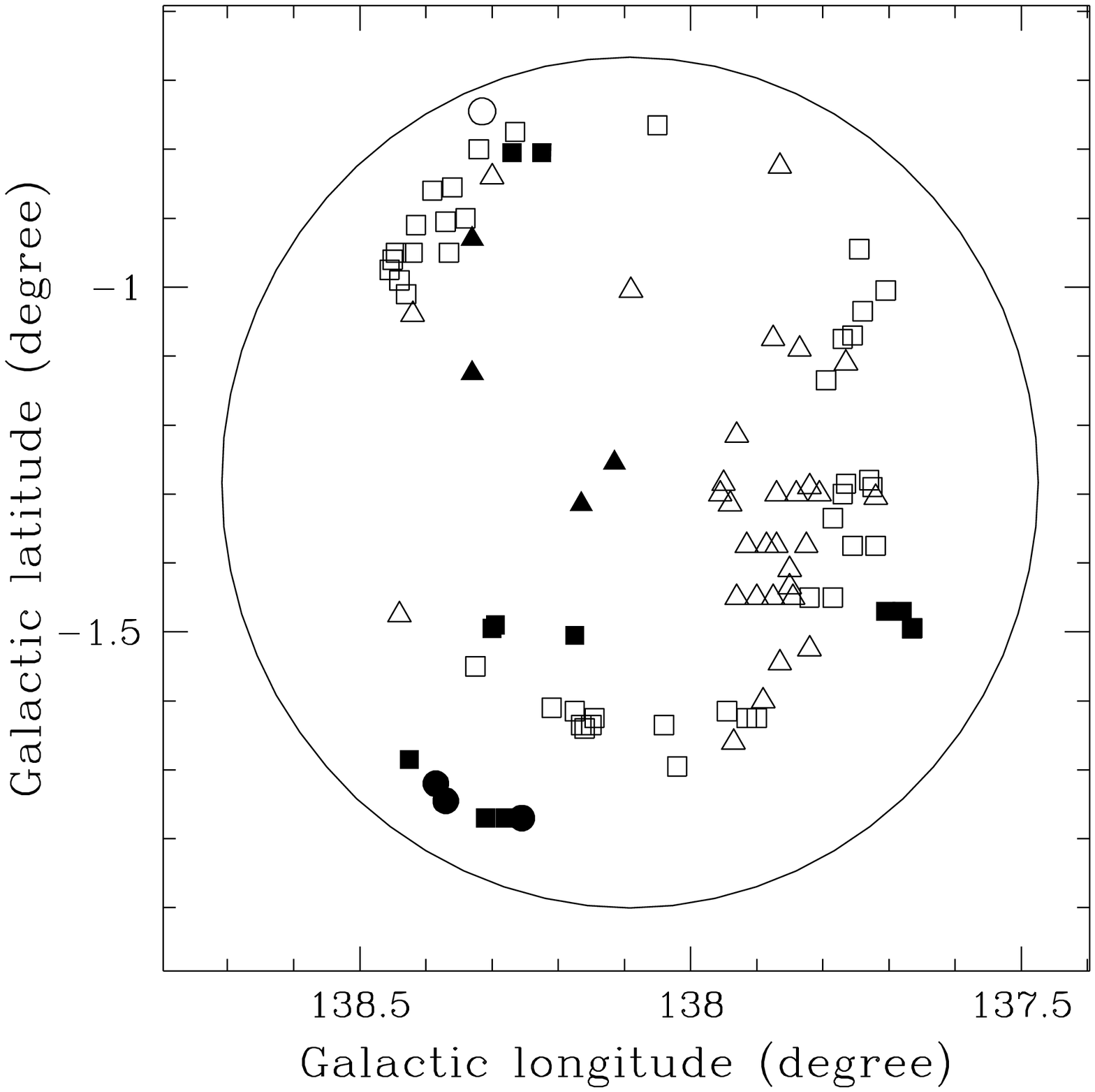}{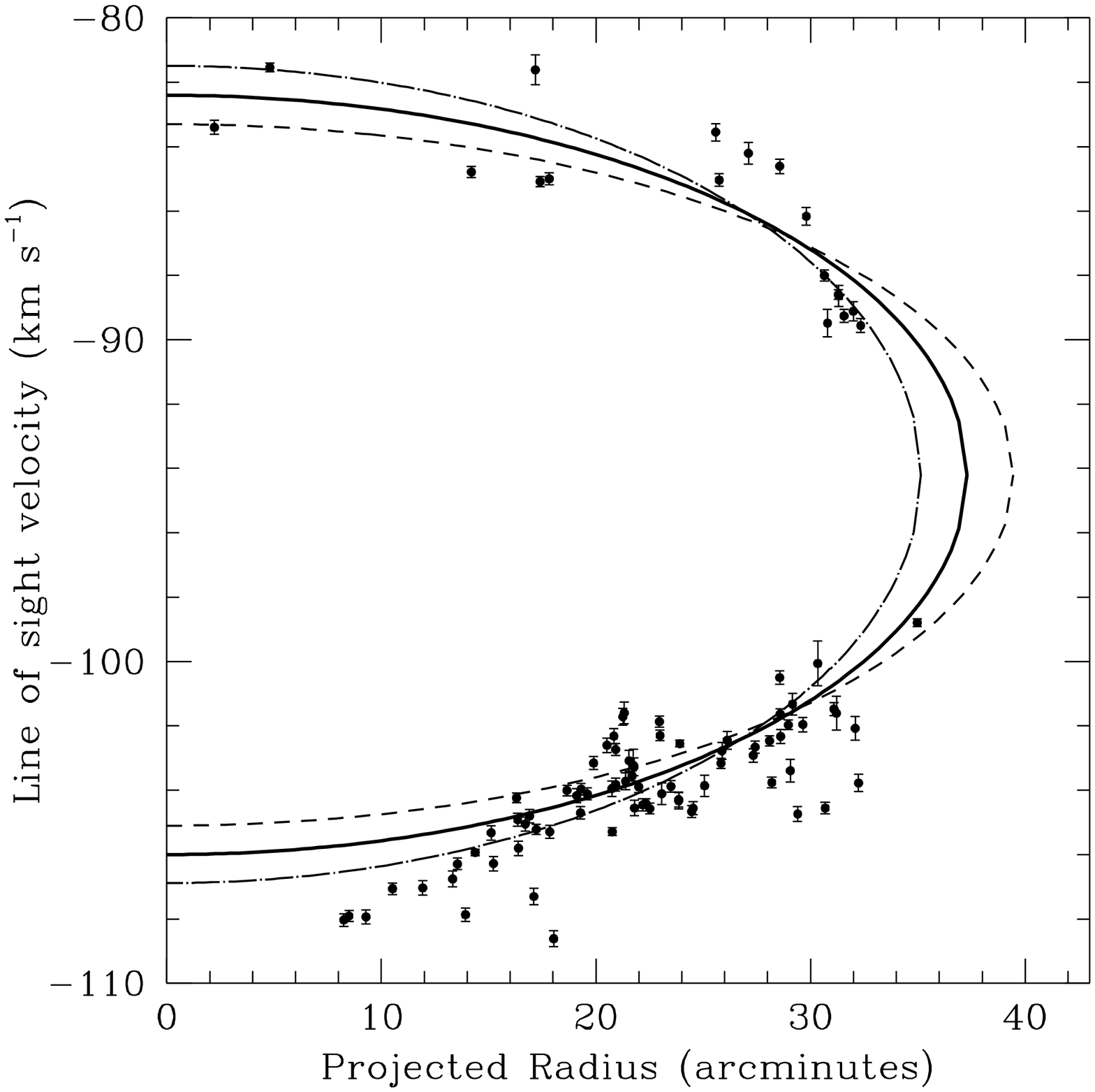}
\caption{{\it Left:\/} Locations of data points used for the fit on
the sky. Symbols: triangles: $|v_{\rm los}-v_{\rm c}|>10\kms$,
squares: $10{\kms} \ge |v_{\rm los}-v_{\rm c}|>5 \kms$, circles: $5
{\kms} \ge |v_{\rm los}-v_{\rm c}|>0 \kms$. Filled symbols represent
the receding (far) side of the shell, open symbols the approaching
(near) side.  {\it Right:\/} Line of sight velocity of shell material
as a function of projected distance from the centre. The curved lines
are the best fit solution of expression (1) (solid line) and the
solutions with 1$\sigma$ perturbations of radius and expansion
velocity (dashed and dot-dashed curves).
\label{fitres-fig}
}
\end{figure}

\begin{table}
\begin{center}
\caption{Properties of \GSH.\label{fitres-tab}}
\begin{tabular}{  l   r    l  } 
\tableline\tableline 
Quantity     &  Value \ \ \ \ \ \ \   & Reference \\ 
\tableline
Centre (Galactic) & $l = 138\degr6\arcmin\ \pm 3\arcmin$ $b = -1\degr15\farcm5\ \pm 3\arcmin$ & Section~\ref{properties-sec} \\
Centre (J2000)  & $\alpha=2^{\rm h}50^{\rm m}35\fs0$ $\delta=+58\degr00\arcmin15\arcsec$ & \\
Central velocity & $-94.2 \pm 0.5\ \kms$ &  Section~\ref{properties-sec}\\
Expansion velocity  & $11.8 \pm 0.9\ \kms$ &  Section~\ref{properties-sec} \\
Heliocentric distance  & $16.6 \pm 2.5\ \rm kpc$  &  Section~\ref{distance-sec} \\
Galactocentric distance  & $24\ \rm kpc$  &  Section~\ref{distance-sec} \\
Radius  & $37\farcm3 \pm 2.2$ &  Section~\ref{properties-sec} \\
        & $180 \pm 10\ \rm pc$ & \\
Mass    & $2 \times 10^5\ \rm M_\sun$ &  Section~\ref{mass-sec} \\ 
Expansion kinetic energy &  $3 \times 10^{43}\ \rm J$  &  Section~\ref{mass-sec} \\
Expansion momentum   &  $2 \times 10^{6}\ \rm M_\sun \kms$   &  Section~\ref{mass-sec} \\
Expansion age        & $4.3 \times 10^{6}\ \rm year$ &  Section~\ref{mass-sec} \\
\tableline
\end{tabular} 

\end{center}
\end{table}

The centre, radius, expansion velocity and central velocity of the
shell were determined by fitting the velocity field of a thin
expanding shell to the data at a number of positions. The line of
sight velocity $v_{\rm los}$ for a thin expanding shell with centre
$(x_{\rm c},y_{\rm c})$, expansion velocity $v_0$, radius $R_{\rm S}$,
and central velocity $v_{\rm c}$, at any point $(x,y)$, with
$(x-x_{\rm c})^2 + (y-y_{\rm c})^2 < R_{\rm S}^2$, is given by
$$
v_{\rm los}=v_{\rm c} \pm v_0 \sqrt{1-{{(x-x_{\rm c})^2 + (y-y_{\rm c})^2}\over R_{\rm S}^2}}
\eqno(1)
$$
The five parameters defining this function can be determined by
fitting a large number of measured points $(x_i,y_i,v_i)$.  In
particular, the expansion velocity of the shell is determined by on
and off-axis data rather than only by emission at the extreme
velocities. Similarly, the radius is determined from points over many
channels rather than from a single channel map.

The data points $(x_i,y_i,v_i)$ may in principle be distributed
anywhere within the radius of the shell, although in practice, there
are some restrictions.  Measurements well within the outer boundary
are required to allow a determination of the expansion velocity.
Measurements near the outer boundary of the shell should be avoided,
because the component of the expansion velocity along the line of
sight vanishes there. Furthermore a degeneracy between the radius and
expansion velocity of the shell exists if all measured points are at
the same distance from the centre.  Data from both sides of the shell
are not required, but more accurate results can be expected if data
from both the approaching and receding side are included.

Line of sight velocities at 94 positions throughout the area of \GSH
were determined by fitting the local line profile with 3 to 5
Gaussians and a second degree polynomial baseline. We are only
concerned with obtaining a good fit to the centroid of the one
component in the line profile associated with the shell. The remaining
components were added to achieve a satisfactory fit to the \HI line
profile.  Only the position and velocity of the Gaussian component
identified with the shell were used. The distribution of the measured
points on the sky is shown in Figure~\ref{fitres-fig}. More data exist
for the approaching side of the shell because this side does not
suffer from confusion with other emission. The results of the fits
(see Table ~\ref{fitres-tab}) did not change by a large amount when
the fit was repeated for five randomly selected subsets containing
only one third of the data.  The spread in the results from the fits
to subsets of the data was used to determine the errors listed in
Table~\ref{fitres-tab}. The results are consistent within the errors
with estimates of the geometric centre and radius defined by the
average and half the difference of the coordinates of opposite sides
of the shell in the channel maps.

Adopting the centre derived from the fit, we plot $v_{\rm los}$ as a
function of projected distance from the centre in order to display the
data and the fit in a 2-dimensional graph. This graph is shown in
Figure~\ref{fitres-fig}. The data points trace the Doppler ellipse
outlined in Figure~\ref{shellxv-fig}. The scatter in the data exceeds
the formal error in the Gaussian fits, and likely reflects departures
from the idealized expansion law assumed in equation (1).  The solid
line represents the best fit to all the data, and the interrupted
lines indicate the errors quoted for the radius and expansion
velocity. Note that a smaller expansion velocity requires a larger
radius to obtain a reasonable fit.  The most extreme negative
velocities are not very well represented by the fit. These points can
be traced to the vertical filament visible in the channel maps around
$-107.8\kms$.

\subsection{Distance}
\label{distance-sec}

The central velocity of \GSH is close to the extreme velocity of
Galactic \HI. This implies a large galactocentric radius for any
choice of Galactic rotation curve that is reasonably flat.  As many
derived properties depend strongly on the adopted distance, a careful
review of the uncertainties involved is essential.

The line-of-sight velocity of a cloud on a circular orbit with radius
$R$ and velocity $V(R)$ around the Galactic centre assuming circular
orbits, is
$$
v_{\rm los} = \Bigl({R_0\over R}V(R)-V(R_0)\Bigr) \sin l  
$$
Here, $R_0$ is the distance of the sun to the Galactic centre, and $l$
is the Galactic longitude of the cloud.  This relation can be solved
for the Galactocentric distance $R$, and hence the heliocentric
kinematic distance $d$. For \GSH, $v_{\rm c}=-94.2\kms$, and
$l=138\degr6\arcmin$, so $R=23.6\ \rm kpc$, and $d=16.6\ \rm kpc$,
assuming $V(R)=220 \kms$ and $R_0 = 8.5 \ \rm kpc$.

If \GSH is not at such a large distance, then it must have a high
anomalous velocity with respect to the surrounding medium. We rule
this out for two reasons. First, the appearance of \GSH on the sky is
nearly perfectly circular. This effectively rules out a velocity
relative to its immediate surroundings which is a large fraction of
the expansion velocity ($11.8 \kms$). Second, the mass of \GSH is
likely dominated by swept-up interstellar gas. Suppose the source of
the shell has a velocity of $100 \kms$ relative to the ambient medium,
and the mass of the ejecta amounts to 10\% of the mass of the shell.
Conservation of the momentum of the ejecta due to the velocity of the
source, implies that the peculiar velocity of the shell would be only
$10 \kms$ in this example.

Of course, this argument does not exclude that the velocity of the
surrounding medium is also different from standard Galactic rotation.
We now consider localized departures from Galactic rotation along the
line of sight, which could create the false impression of a large
distance, and departures from standard Galactic rotation in the outer
Galaxy.

Large-scale streaming motions in galaxies are associated with bars or
spiral arms. Streaming motions associated with spiral arms are the
only relevant factor outside the solar circle. The spiral shock model
of \citet{roberts72} for the longitude range $l=130\degr$ to
$l=140\degr$ allows a minimum line of sight velocity of approximately
$-55 \kms$, which is some $20 \kms$ blueshifted from standard Galactic
rotation. The velocity of \GSH is not allowed by the density wave
model for the Perseus arm by a margin of $40 \kms$.  Also, absorption
of radio continuum emission towards the W3/W4 \HII regions is not
observed beyond $v_{\rm LSR}=-50 \kms$ \citep{normandeau99}.  We
therefore conclude that the velocity of \GSH implies that it is
located well beyond the Perseus arm.

The validity of standard Galactic rotation at large radii is
uncertain.  The rotation velocity of most spiral galaxies remains
nearly constant to the outermost point that can be measured, and the
same is true for the Galaxy to a distance $2 R_0$ from the centre
\citep{merrifield92}. By comparison, the kinematic distance of \GSH
implies it is $2.8 R_0$ from the Galactic centre. The issue concerning
the distance to \GSH is not whether the rotation curve of the Galaxy
declines at large distances, but rather whether the assumption that
the gas follows circular orbits is valid. The magnitude of this effect
may be estimated from the $l \rightarrow -l$ asymmetry of the extreme
velocity of \HI in the Galaxy \citep{kuijken94}. The magnitude of this
asymmetry is at most $20\kms$, and the sign implies a more negative
velocity in the second quadrant. We estimate the effect of
non-circular orbits by adding a radial expansion term of $-10 \kms$ to
the velocity $V(R)$ and recalculating the kinematic distance. The
result is a distance which is $3\ \rm kpc$ or 20\% smaller than the
distance derived assuming circular orbits.

In this paper we therefore adopt a distance of $16.6\ \rm kpc$ for
\GSH, keeping in mind that a systematic effect of the order of 20\%
may exist due to non-circular orbits in the outer Galaxy.

\subsection{Environment}
\label{env-sec}

The large distance of \GSH places it in the far outer Galaxy. The
Galactic latitude of its centre corresponds with a distance 370 pc
south of the Galactic plane. The actual centre of the \HI layer cannot
be determined with a high accuracy for a particular longitude. From
Figure 60b in \citet{burton91} it is estimated that the centre of the
\HI layer is between 0.5 and 1 kpc toward positive Galactic latitude
in the vicinity of \GSH.  The half-width thickness of the Galactic \HI
layer at 20 kpc from the centre is approximately 1.5 kpc
\citep{merrifield92}. Therefore, \GSH is not exceptionally far from
the local midplane, and its diameter is about a quarter of the
thickness of the disk at half maximum. The side of \GSH in the
direction of the Galactic equator is a factor two brighter than the
side facing the opposite direction, which is consistent with a density
gradient perpendicular to the plane.

Adjacent to \GSH at lower longitude an \HI complex is observed that
may have some connection with \GSH. The \HI complex displays a
peculiar velocity gradient in the sense that the velocity is more
negative at a larger distance from the Galactic equator. For a disk in
which the rotation speed does not change with distance from the
midplane, the velocity is expected to become less negative by a factor
$\cos(b)$, which is negligible over the latitude range
considered. Although this \HI complex may be physically related to
\GSH, it may also be a coincidence because a small velocity interval
corresponds with a large distance along the line of sight in this part
of the Galaxy.

The gas density in the vicinity of \GSH was derived according to the
relation $n_{\rm H}=N_{HI}/\Delta d)$, with $N_{HI}$ the \HI column
density obtained by integrating the brightness temperature over a
narrow velocity interval $\Delta v$ centered on the central velocity
of the shell. $\Delta d$ is the length of the line of sight
corresponding to $\Delta v$ and Galactic rotation. The brightness
temperature averaged over an area that covers the latitude range of
\GSH on both sides in longitude is $T_{\rm b}= 13\ \rm K$, and does
not vary significantly with velocity for $\Delta v < 10 \kms$. The
resulting hydrogen density is $n_{\rm H}=0.02 \ \rm cm^{-3}$, which is
fairly independent of the particular choice of $\Delta v$.  This
density is within a factor 2 of the azimuthally averaged \HI density
$n_{\rm H}=0.04$ at $R=2.35 R_0$ in Figure~8 of \citet{merrifield92}.
A consequence of the steep slope of the velocity-distance relation for
the longitude and velocity of \GSH is that the derived density is the
average across a significant line of sight: $\Delta d = 4.7\ \rm kpc$
for $\Delta v = 10 \kms$. This will be discussed further in
Section~\ref{mass-sec}.

\subsection{Substructure}
\label{substruc-sec}

At the resolution of 1 arcminute a great deal of substructure is
observed.  At the most extreme velocity ($v_{\rm LSR}=-110 \kms$) a
prominent filament is seen in the channel maps. Such condensations and
filaments may form as a result of the Rayleigh-Taylor instability if a
dense shell is expanding due to the pressure of the tenuous hot
interior, and thermal instabilities because the cooling increases with
density.  A number of compact spots can be seen in
Figure~\ref{chanmap-fig}, e.g. at
$(l,b,v)=(138\fdg125,-0\fdg695,-106.2)$.

At $(l,b,v)=(137\fdg79,-0\fdg81,-99.6)$ and adjacent channels, a
conspicuously empty, $7\farcm2 \times 4\farcm2$ elliptical area is
visible in the rim of the shell.  Over the velocity $-90 \kms > v_{\rm
LSR} > -110 \kms$, the brightness temperature remains below 15 K in
this area. This is consistent with the intensity of the surroundings
of the shell, and about half the intensity of the bright edge
surrounding this area. The isophotes of the hole are elliptical, with
major axis in position angle $36\degr$, and axial ratio 0.59. The
projected distance of the hole from the centre of \GSH is 0.895 times
the radius of the shell in position angle $-33\degr$. A circular hole
in the shell seen in projection at the same location, would appear as
an elliptical hole with its major axis in position angle $33\degr$,
and axial ratio 0.45. Taking into account the uncertainties in the
positions, the observed shape of the hole is consistent with a nearly
circular hole in \GSH. At the distance of 16.6 kpc, the diameter of
this hole is 35 pc. We refer to this hole as Cavity 1.

A smaller, but very distinct empty region is observed at
$(l,b,v)=(137\fdg54,-1\fdg05,-101.2)$, and the adjacent channel
maps. The central brightness temperature remains consistent with the
background, which is a factor 3 below its immediate surroundings. We
refer to this feature as Cavity 2.

\subsection{Spin temperature}
\label{temp-sec}

A weak but distinct absorption feature is observed against a $T_{\rm
b}=82.8\ \rm K$ background continuum source in the direction
$(l,b)=(137\fdg93,-1\fdg27)$. Figure~\ref{profiles-fig}~A shows the
absorption profile in this direction. We determine the peak brightness
temperature by fitting a Gaussian and a constant offset to the data in
a specific velocity interval. The sum of the Gaussian and the offset
is a relatively accurate estimate of the brightness temperature at
$v_{\rm LSR}=-107 \kms$, despite the modest signal to noise ratio of
typically 7, because it uses information from more than one channel.
This was done for eight positions, $\pm 3\arcmin$ offset in longitude
or latitude, from the position of the continuum source. The fitted
offset in intensity contributes typically less than 20\% to the total
intensity.  The mean brightness temperature at the eight off-positions
is $T_{\rm b,off}=18.6\ \pm\ 1.3\ \rm K$. The error indicates the rms
variations, divided by $\sqrt 8$, and is comparable to the formal
error in the fit of a single spectrum. At the position of the
continuum source the brightness temperature is $T_{\rm b, on}=11.9\ \pm\
1.5\ \rm K$, which is less than any of the eight surrounding
positions. These numbers imply a spin temperature between $173\ \rm K$
and $367\ \rm K$, for $1 \sigma$ variation of $T_{\rm b,on}$ and
$T_{\rm b,off}$, with a most likely value of $T_{\rm s}=230\ \rm K$. The optical
depth is 0.084. Although velocity crowding can place unrelated gas at
the same velocity, the fact that we see a narrow absorption feature at
this velocity argues against this possibility.  We therefore conclude
that the spin temperature derived here applies to \HI in \GSH.

An upper limit to the kinetic temperature can be derived from the
width of the line profiles. The average \HI velocity dispersion is
$1.9 \kms$ for the eight off-positions, with scatter consistent with
the formal error of $0.4\kms$ from the fit. Fits at other locations
in the shell give a similar result.  An upper limit $T_{\rm kin}< 430 \
\rm K$ to the kinetic temperature is found by assuming that the
observed linewidth is completely due to thermal broadening. This upper
limit is consistent with the low spin temperature derived above, and
emphasizes that the temperature in the shell is not far from the
temperature of warm \HI in the interstellar medium.

\subsection{Mass, energy, momentum}
\label{mass-sec}

The relatively smooth \HI background visible in
Figure~\ref{chanmap-fig} was subtracted by applying a median filtering
procedure outlined in Section~\ref{cont-sec} to the \HI channel maps,
with one difference.  Instead of applying a cutoff in intensity to
exclude bright point sources, the brightest parts of the shell were
masked manually in each channel map.  Inspection of the line profiles
next to the brighter parts of \GSH showed that the zero level in the
residual data cube is well defined. Some uncertainty is unavoidable
because emission of \GSH is blended with the \HI cloud on the right
hand side of \GSH in the velocity interval $-94.6 \kms$ to $-101.2
\kms$, around longitude $137\fdg5$.

Summation of the background subtracted data cube between $-82.3 \kms$
and $-115.2 \kms$ results in a total \HI mass $M_{\rm HI}=1.8 \times
10^5\ \rm M_\sun$ within a circular area defined by the outline of
\GSH. The area bounded by the limits $l<137\fdg75$, $b>-1\fdg45$, and
the outer radius of \GSH is affected by confusion with the adjacent
cloud (see Figure~\ref{chanmap-fig}). The \HI mass in this area is
$4.2 \times 10^4\ \rm M_\sun$, or 23\% of the total. In order to
correct for the presence of the cloud, we assume that the \HI mass in
the area affected by confusion is equally distributed between the
shell and the cloud. With this correction, we find $M_{\rm HI}=1.6
\times 10^5\ \rm M_\sun$.  The atomic hydrogen mass is an order of
magnitude larger than the molecular mass discussed in
Section~\ref{CO-sec}. Taking into account primordial Helium, we find
the total mass of \GSH is $M_{\rm tot}=1.3 M_{\rm HI}$.  The kinetic
energy associated with the expansion of the shell then follows from
$E_{\rm kin}= {1 \over 2} M_{\rm tot} v_{\rm exp}^2$ and the momentum
from $P_{\rm exp}= M_{\rm tot} v_{\rm exp}$. Numerical values for
these quantities are listed in Table~\ref{fitres-tab}.

The hydrogen mass in the volume of the shell assuming the density
$n_{\rm H}=0.02\ \rm cm^{-3}$ (Section~\ref{env-sec}) is $1.2 \times
10^{4}\ \rm M_\sun$, an order of magnitude smaller than the \HI mass
of the shell. This difference cannot be the result of an error in the
distance. The swept-up mass is proportional to the volume, therefore
the third power of distance. The HI mass derived from the lineflux
is proportional to the square of the distance. Therefore, a ten times
larger distance would have to be assumed to obtain consistent results.
The discrepancy can be resolved if the environment of \GSH has a
higher density than the Galactic average at the Galactocentric radius
of \GSH.  The density found in Section~\ref{env-sec} is an average
over a long line of sight. If the swept-up mass and the mass of the
shell are equal, the density is $n_{\rm H}=0.27\ \rm cm^{-3}$, seven
times the average midplane density found by \citet{merrifield92}. This
estimate of the ambient density is preferred over the density derived
in Section~\ref{env-sec}, but the uncertainty in the density is at
least a factor 2.

The expansion age of \GSH is $\tau_{\rm exp} = \alpha R/v_{\rm exp}$,
assuming the expansion law $R\sim t^\alpha$.  We assume $\alpha={2
\over 7}$. This is applicable to a shell expanding from the pressure
of a hot interior, such as an old supernova remnant \citep{mckee77}.
The expansion age thus becomes $\tau_{\rm exp} = 4.3\ \rm Myr$. The
precise value of $\alpha$ is of less importance given the
uncertainties, but we note that we somewhat underestimate the age in
the likely event that the expansion law was steeper in the past.

\subsection{Radio continuum}
\label{cont-sec}

\begin{figure}
\caption{ Gray scale representation of the 408 MHz continuum image.
Gray scales are linear and range from 50 K (white) to 65 K (black).
The location of the \HI shell is indicated by the white contour, of
the column density $2\times 10^{20}\ \rm cm^{-2}$, obtained by
integration over the velocity range $-98\ \kms$ to $-107\ \kms$.  A
$+$ marks the expansion centre of \GSH determined in
Section~\ref{properties-sec}. The square indicates the field of view
of Figure~\ref{C21-fig}.
\label{C74-fig}}
\end{figure}

\begin{figure}
\caption{ Detail of \GSH displaying the area around Cavity 1 indicated
by the box in Figure~\ref{C74-fig}. Gray scales represent \HI column
density, integrated over the velocity range $-98 \kms$ to $-107
\kms$. The range of gray scales is $0.6 \times 10^{20}\ \rm cm^{-2}$
(white) to $8.0 \times 10^{20}\ \rm cm^{-2}$ (black). White contours
give the CO brightness temperature integrated over the same velocity
range. Contour levels for the CO map are 2 (5$\sigma$), 3, 4, 5 $\rm
K\kms$. The 21-cm continuum intensity is shown as black contours at
$T_{\rm b} = $ $0.14$ ($2\sigma$), $0.28$, $0.42$, ...  K above the
mean background level of $5.13$ K. The $+$ indicates the position of
the B1 star studied by \citet{degeus93}. The peaks in the H$\alpha$
emission are indicated with $\times$.
\label{C21-fig}}
\end{figure}

Figure~\ref{C74-fig} shows the DRAO 74-cm continuum map of the area of
\GSH.  The gradient in the background intensity in
Figure~\ref{C74-fig} is due to the Galactic background emission that
decreases with distance from the Galactic equator.  There is no
indication of enhanced emission, on top of the Galactic background,
associated with \GSH.  In order to derive an upper limit to the
continuum flux of \GSH, we subtracted the smooth Galactic background
emission in the following way.

First, point sources brighter than approximately $T_{\rm b}=20\ \rm K$
above the diffuse background were excluded from the analysis by
clipping on intensity. The clipped image was subjected to a spatial
median filter with dimensions $6\arcmin\ \times\ 90\arcmin$ in
latitude and longitude, in two steps. In the first step, the median
filtered image was subtracted from the original image in order to
obtain a first order residual image of structure on scales much
smaller than the size of the median filter. This first-order residual
image was clipped at the $T_{\rm b}=4\ \rm K$ level, setting all
pixels below this level to 0. Only individual point sources exceeded
the $4\ \rm K$ level. The clipped first-order residual image was then
subtracted from the original image.

In the second step, the median filter was applied to the point source
subtracted image.  This second pass resulted in our final estimate of
the smooth background.  The residual image after subtraction of the
final smooth background image, has a well-defined zero level. The
elongated shape of the median filter allows effective subtraction of
structure that is extended in Galactic longitude, while retaining
elongated structure oriented at a large angle with respect to the
Galactic plane. Negative residuals are less than 7\% (typically 2\%)
of the smooth background emission in the vicinity of \GSH. The
background subtraction procedure was tested in another area of the
CGPS, where continuum shells of the same angular size as \GSH were
detected. It was confirmed that a shell structure of the size of \GSH
is effectively separated from the background by this procedure. This
method, then, effectively allows us to identify a shell in continuum,
if it exists. No shell was observed above the noise.

The sensitivity of the DRAO 408 MHz map is limited by confusion of
faint background sources. Excluding individually visible point
sources, the r.m.s.  fluctuations in the residual image are $\Delta
T_{\rm b} = 0.86\ \rm K$.  Due to the crowding of the many resolved
point sources, it is not feasible to integrate the flux over the area
of \GSH. Instead we adopt the r.m.s.  fluctuations of the background
as a safe upper limit to the surface brightness. An object with the
size of \GSH and brightness temperature $T_{\rm b} = 0.86\ \rm K$ ($1
\sigma$) would have been recognized in the residual map without
difficulty. The number of independent beams over the area of \GSH
should not be used to quantify the statistical significance of this
upper limit, because it does not take into account the many points
sources in the field.

The 408 MHz surface brightness is then bounded by $\Sigma_{408}< 4
\times 10^{-23}\ \rm W\ m^{-2}\ Hz^{-1}\ sr^{-1}$. This upper limit is
significant, because it is an order of magnitude below the surface
brightness of known old supernova remnants \citep{landecker90}.
An upper limit to the 408 MHz flux density can be derived from the
upper limit to the surface brightness. There is some uncertainty in
the choice of the area to which the surface brightness limit
applies. A likely possibility is that a radio continuum counterpart of
\GSH would have a shell morphology similar to the HI. In this case,
the upper limit to the surface brightness is really a limit to the
surface brightness of the bright outer rim of such a continuum
shell. Assuming a ring with outer radius $37\farcm3$, inner radius
$30\arcmin$, and a uniform surface brightness equal to the upper limit
translates in an upper limit of $0.57\ \rm Jy$ to the flux density at
408 MHz.

Figure~\ref{C21-fig} shows the 1420 MHz continuum image of the area
indicated by the square in Figure~\ref{C74-fig}. An extended continuum
source is seen in projection on the edge of Cavity 1.  When rotated by
20 degrees counter clockwise, Figure~\ref{C21-fig} can be compared
directly with Figure~3 in \citet{degeus93}.  Radio continuum emission
is observed at the location of the two maxima in H$\alpha$ intensity
in the map of \citet{degeus93}, as indicated in
Figure~\ref{C21-fig}. In fact, the \HII region is extended and covers
the same area on the sky as the faint resolved 21-cm continuum
emission.  Based on the correlation with H$\alpha$ and the absence of
a counterpart at 74 cm, the continuum emission may well be thermal
emission from the \HII region.  The velocity of the \HII region as
determined by \citet{degeus93}, is $-101 \kms$. The \HI that can be
associated with this \HII region and the adjacent molecular cloud
(Section~\ref{CO-sec}) is part of the edge of Cavity 1, which is a
hole in \GSH (Section~\ref{substruc-sec}).

\subsection{Molecular gas}
\label{CO-sec}

\begin{figure}
\plotone{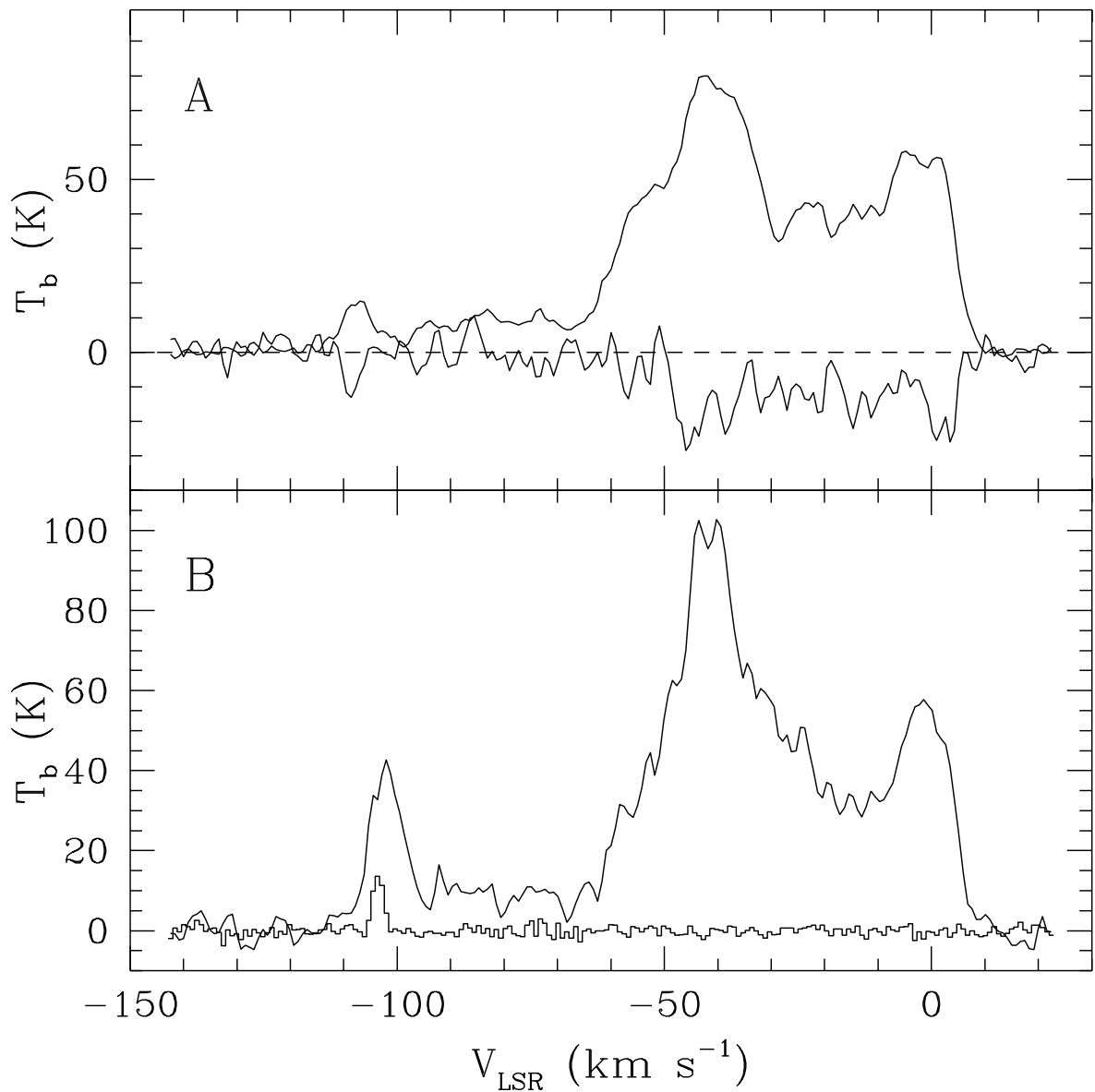}
\caption{ A: Absorption of Galactic \HI against the continuum source
at $(l,b)=(137\fdg93,-1\fdg27)$. The smooth emission spectrum is the
average of eight positions $3\arcmin$ offset in $l$ and $b$. The
negative curve is the difference between the spectrum at the position
of the continuum source and the average spectrum of the
surroundings. Note the absorption feature at $v_{\rm LSR}=-108 \kms$,
which is associated with \GSH. B: CO emission line (histogram) and \HI
emission in the direction of the CO cloud.
\label{profiles-fig}
}
\end{figure}

The FCRAO CO data show a small molecular cloud (number 2 in
Digel et~al. 1994) at $(l,b,v)=(137\fdg76,-0\fdg97,-103.6)$, with peak
brightness temperature $T_{\rm b}= 2.9\ \rm K$, and velocity dispersion
$\sigma_{\rm CO}= 0.94 \kms$. \citet{degeus93} found an \HII region
with the same velocity as the molecular cloud and a candidate ionizing
B star with a photometric distance between 17.5 and 32 kpc. The relative 
locations of the molecular cloud, the \HII region and the ionizing star are 
indicated in Figure~\ref{C21-fig}. 

Figure~\ref{profiles-fig}~B shows the CO and \HI spectra at this
position.  \HI emission at $v_{\rm LSR}=-101.9 \kms$ is associated
with \GSH. The \HI and CO profiles can be compared directly thanks to
the arcminute resolution and zero spacing information in the DRAO \HI
data.  The coincidence of these two features in position {\it and}
velocity is highly unlikely to be a chance alignment. We therefore
associate the CO cloud with the approaching side of the \HI
shell. Note that this association of the CO cloud with \GSH means its
distance is $16.6\ \rm kpc$ (Section~\ref{distance-sec}), considerably
smaller than the kinematic distance of 21 kpc adopted by
\citet{digel94}. It is noteworthy that cloud 2 is the only one of 11
clouds in the sample of \citet{digel94} for which the mass derived
from the CO flux exceeds the virial mass.  The smaller distance that
we derived here, decreases this discrepancy significantly, because the
virial mass scales only linearly with distance.

In this interpretation, this molecular cloud is only marginally more
distant than the other molecular clouds in the sample of
\citet{digel94}, and only marginally inconsistent with the photometric
distance of the B star found by \citet{degeus93}. \citet{digel94}
found the mass of this molecular cloud to be $2.3 \times 10^4\ \rm
M_\sun$, assuming a distance of 16.6 kpc. The bright northern rim of
the \GSH was targeted by \citet{digel94}, but no CO emission was found
there. \citet{digel94} identified the pointsource IRAS 02450+5816 with their 
cloud 2, but there is no indication of a shell structure in the IRAS maps.

\section{Discussion}

\begin{table}
\caption{ Properties of \GSH compared with model predictions derived
from the scaling relations in \citet{thornton98}. The energy released
by the supernova was assumed to be $10^{44}\ \rm J$. The properties of
\GSH are given on the last line.\label{thornton-tab}} \begin{tabular}{  c  c   c   c  c  c  } 
\tableline\tableline
 $n_{\rm H}$          &$\rm log([Z/Z_\sun])$  &  $R_{\rm S}$     & $v_0$  &${\rm log}(M_{\rm tot}/\rm M_\sun)$ \\ 
($\rm cm^{-3}$)   &                        &  (pc)    & ($\rm km\ s^{-1}$) &   \\ 
\tableline
 0.03 & $-1.0$ &  242.44  &  14.36 &   4.76 \\
 0.03 & $ 0.0$ &  192.57  &  11.67 &   4.49 \\
 0.09 & $-1.0$ &  152.83  &  14.20 &   4.64 \\
 0.09 & $ 0.0$ &  121.40  &  11.55 &   4.37 \\
 0.27 & $-1.0$ &   96.34  &  14.05 &   4.53 \\
 0.27 & $ 0.0$ &   76.53  &  11.42 &   4.26 \\
\tableline
 0.27 &  -1.0\tablenotemark{a}  &  180$\pm$10 & 11.8$\pm$0.9 &5.3$\pm$0.3 \\ 
\tableline 
\end{tabular} 
\tablenotetext{a}{Metallicity estimated from the Galactic oxygen abundance gradient
\citep{smartt97}.} 

\end{table}

\subsection{Source of the shell}
\label{source-sec}

The kinematic distance to \GSH is the basis for our interpretation of
the data presented in this paper. Summarizing the arguments given in
Section~\ref{distance-sec}, we conclude that systematic effects
between $10 \kms$ and $20 \kms$ may exist, but these do not seriously
affect our conclusions. We estimate an error in the distance of 20\%,
similar to the assessment of \citet{digel94} for their molecular
clouds.  The association of the \HII region found by \citet{degeus93}
and a photometric distance of its probable ionizing star with \GSH
supports the large distance derived in Section~\ref{distance-sec}.
The conclusion that \GSH is in the far outer Galaxy does not depend on
the exact shape of the rotation curve in the outer Galaxy.  The
distance of \GSH is therefore sufficiently secure to allow a
meaningful comparison with other shells of known origin.

The size of \GSH is similar to small superbubbles, such as the
Orion-Eridanus bubble \citep{brown95}. However, the kinetic energy of
the expanding shell ($3 \times 10^{43}\ \rm J$) is consistent with the
$10^{44}\ \rm J$ released in a single supernova explosion
\citep{leitherer92}. Although the expansion energy of \GSH is a factor
3 smaller, this is well within the uncertainty in the energy
release of a supernova, and which fraction of this energy is in the
form of kinetic energy of the ejecta. A recent discussion was given in
\citet{thornton98}. The energy released in a stellar wind can be of
the same order of magnitude as the kinetic energy of the ejecta of a
supernova explosion. We first explore what star could be the source of
\GSH assuming the shell is the result of only a stellar wind.

\citet{normandeau00} discuss the possible sources of the \HI shell
G132.6$-$0.7$-$25.3 (radius 33 pc). This shell may be a bubble blown
by the stellar wind of the B1 Ia star BD$60\degr 447$, but the
expansion age of this shell is less than the age of the star.
\citet{normandeau00} conclude that G132.6$-$0.7$-$25.3 can also be an
old supernova remnant. \citet{bransford99} found shells with radii up
to 80 pc in a survey of Wolf-Rayet stars in M31.  Although \GSH is
more than twice the size of the largest of these shells, it cannot be
excluded that this is the result of a lower ambient density.  The most
significant parameters for the interpretation of \GSH as a wind blown
bubble are the expansion kinetic energy and the age of the shell. The
expansion kinetic energy of \GSH is 30 times larger than that of
G132.6$-$0.7$-$25.3 \citep{normandeau00}. Note that
\citet{normandeau00} found only one star that could have been the
origin of G132.6$-$0.7$-$25.3 by means of its stellar wind.

The mass loss rate from a stellar wind increases rapidly with
luminosity. The longer lifetime of a star with lower mass compensates
only partly for the much smaller mass loss rate. The mass loss rate
also depends on the metallicity Z.  \citet{nugis00} find that the mass
loss rate scales as $Z^{m}$, $m=0.5$. \cite{vink01} predict $m=0.64$
to $m=0.69$, and argue that oxygen is the relevant trace element for
$Z$ at low metallicity. At 18 kpc from the Galactic center, the
abundance of oxygen relative to hydrogen is a factor 10 below the
solar value \citep{smartt97}. The velocity of the stellar wind is also
smaller at low metallicity \citep{nugis00}. We conclude that the
energy released in the form of a stellar wind decreases by a factor
$\sim$3 if the metallicity decreases with a factor 10, for stars with
the same luminosity.

The relation between stellar luminosity and the energy released per
second in the stellar wind was defined from the sample of O and B
stars of \citet{lamers99}. The lifetime of these stars was related to
luminosity at mid-life using the stellar evolution models for solar
metallicity of \citet{bressan93}. From this, a relation was found
between luminosity and the total energy released in the stellar wind
during the lifetime of the star. The expansion energy of \GSH requires
a star with luminosity ${\rm log}(L/L_\sun)=5.7$ at solar
metallicity. Assuming the metallicity is 0.1 solar, we find ${\rm
log}(L/L_\sun)=5.9$. This luminosity implies a star with main sequence
mass $60\ \rm M_\sun$, lifetime 4.1 Myr \citep{bressan93}, spectral
type on the main sequence approximately O5. The same result can be
obtained from \citet{abbot82}. It was implicitly assumed that all of
the kinetic energy of the stellar wind is available for the expansion
of the shell. It is more likely that some fraction is converted into
thermal energy as the stellar wind and the swept-up interstellar
medium pass through a shock. In this case a stronger stellar wind
(i.e. a more massive star with a shorter lifetime) would be
required. The age of \GSH is 4.3 Myr assuming an expansion law $R \sim
t^{2/7}$ (Section~\ref{properties-sec}), similar to the lifetime of a
$60\ \rm M_\sun$ star. It is therefore difficult to provide the
expansion energy of \GSH in the form of a stellar wind by a star with
a lifetime as long as the age of \GSH.

Are the present data consistent with the existence of such a massive
star? If the star still exists, thermal emission of an \HII region
associated with the star should be visible in the 1420 MHz continuum
map, which shows only the small \HII region found by \citet{degeus93}
(Figure~\ref{C21-fig}).  The effective temperature is lower at a later
stage of evolution of the star, but only for a brief period of
time. In the models of \citet{bressan93} the effective temperature of
a star with initial mass $60\ \rm M_\sun$ is below $3 \times 10^4\ \rm
K$ for only 4000 years (0.1\% of the lifetime of the star). Assuming
an O5V star and the extinction $A_{\rm V}=2.7\ \rm mag$ found by
\citet{degeus93}, we estimate the star should be magnitude $V=13$,
observed color $U-B=-0.5$. With this brightness and color, the star
would be an obvious target for the search for blue stars by
\citet{lanning77}. There is no entry in his list near the center of
\GSH, although the objects 7 ($m_{\rm B}\approx 20$) and 54 ($m_{\rm
B}\approx 12.7$) are within the field covered by the channel maps in
Figure~\ref{chanmap-fig} (but outside the perimeter of the shell).

We conclude that the existence of a star with the mass required by the
interpretation of \GSH as a stellar wind bubble is unlikely. The radio
continuum maps at 408 MHz and 1420 MHz also show no evidence for a
supernova remnant inside \GSH. If the shell is a supernova remnant,
the problems raised by the interpretation of \GSH as a stellar wind
bubble are resolved. In particular, a less massive star is required
than in the stellar wind case.

\begin{figure}
\plotone{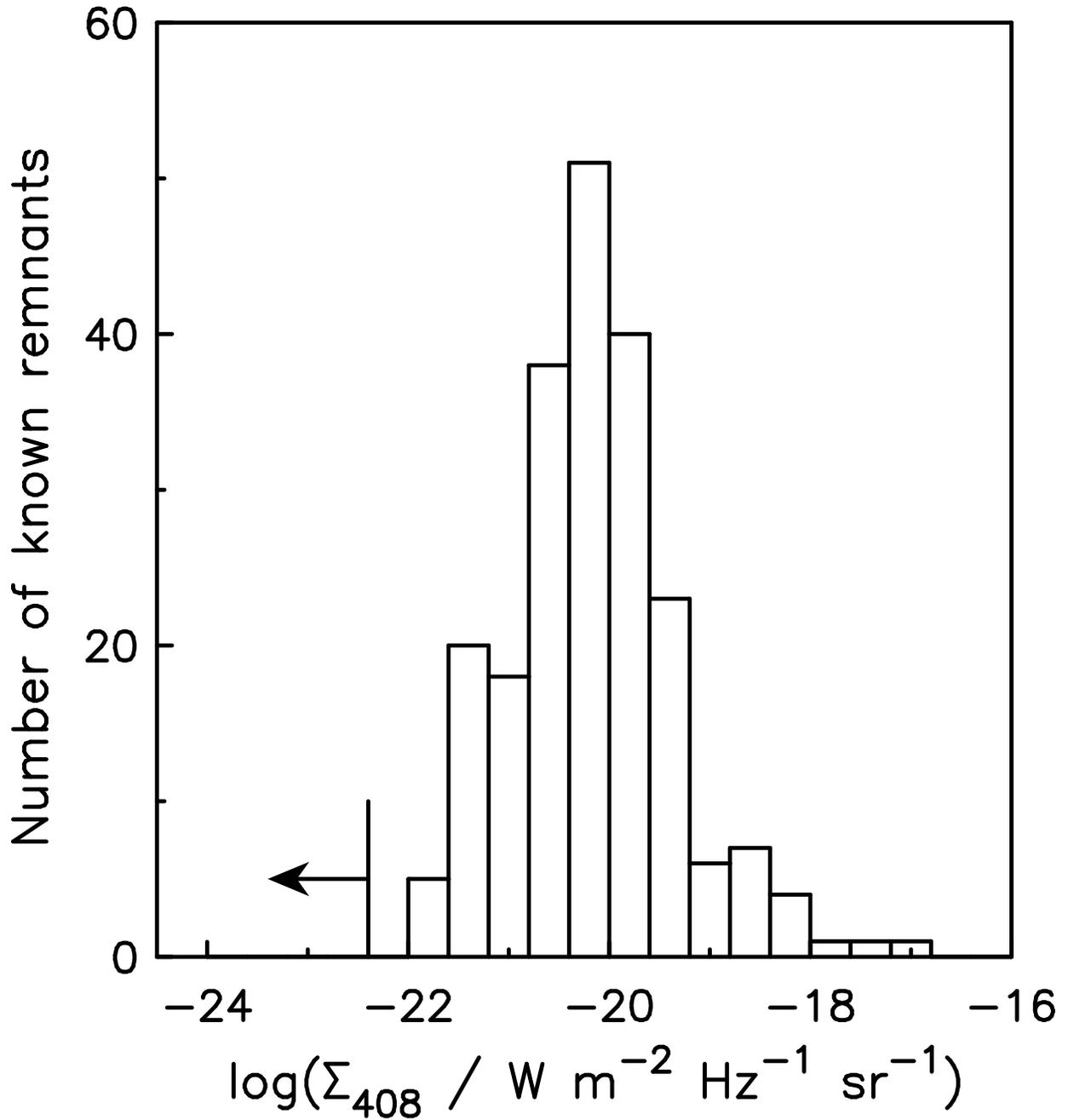}
\caption{ Histogram of the number of known supernova remnants as a function of 
surface brightness at 408 MHz. The arrow indicates the upper limit for 
\GSH derived in Section~\ref{cont-sec}.
\label{SNRhist-fig}}
\end{figure}

\GSH is larger and older than known supernova remnants.  It is
illustrative to compare \GSH with supernova remnants in the solar
neighbourhood.  The Cygnus Loop has a radius of 19 pc and age $1.4
\times 10^4\ \rm yr$ \citep{levenson98}. The Cygnus Loop is therefore
of the same size as Cavity 1 (Figure~\ref{C21-fig}). \GSH is larger
and older than known supernova remnants. Supernova remnants that are
considered to be very old include G55.0+0.3 with radius 70 pc, age 1
Myr \citep{matthews98}, G65.2$+$0.6 with unknown distance but has a
low radio surface brightness, \citep{landecker90}, G106.3$+$2.7 with
radius 70 pc, age 1.3 Myr \citep{pineault00}, and G166.2$+$2.5 with
radius 87 pc, age \citep{routledge86}.  Among the older remnants would
also be radio loop I, with radius 60 pc and expansion velocity $25
\kms$ \citep{heiles84} yielding an age of 0.7 Myr, assuming this is a
single supernova remnant. It is usually difficult to estimate the
distance of large shells that can be old supernova remnants, and to
show that such a shell is the result of a single supernova. Distance
ambiguities can be eliminated when looking at nearby galaxies. The
largest supernova remnant in the optical spectral line survey of M31
by \citet{blair81} is BA~490, with radius 90 pc. Interestingly, BA~490
is located 19 kpc from the centre of M31.  \citet{gordon99} studied a
radio selected sample of supernova remnants in M33, eliminating
distance ambiguities. The largest supernova remnants in their sample
have radii of 50 pc, but these authors acknowledge a selection effect
against the oldest supernova remnants. A direct comparison of \GSH
with known supernova remnants is therefore inconclusive because
cataloged supernova remnants are significantly younger, smaller and
expand faster.  This is likely a selection effect because supernova
remnants are usually identified in radio or X-ray surveys. As the
remnant ages, the radio surface brightness declines and the
temperature of the interior drops.  From this it cannot be excluded
that an older supernova remnant would be similar to \GSH, if it
managed to survive long enough. In particular, the fact that no radio
continuum shell was detected does not disprove the possibility that
\GSH could be a very old supernova remnant.

Figure~\ref{SNRhist-fig} shows the number of known Galactic supernova
remnants as a function of radio surface brightness. This histogram is
based on the most recent (September 1998) version of the catalogue of
\citet{green84,green88} available through the CADC. The surface
brightness at 408 MHz was calculated following \citet{green91},
adopting a spectral index of 0.5 if no spectral index was listed. Six
objects out of 220 were excluded because no radio flux was listed.
The peak of the histogram corresponds with the completeness limit of
radio surveys, as demonstrated in Figure~3 of \citet{green91}.

It is often assumed that an expanding shell dissolves into the ambient
medium if its expansion velocity becomes similar to the random
velocity of clouds, which is of the order of $10 \kms$.  Clearly this
is the case for \GSH, but it is still a well-defined shell. The
residual velocities in Figure~\ref{fitres-fig} are of the order of $3
\kms$. It is not likely that the velocity dispersion in the outer
Galaxy is significantly smaller than $10 \kms$.  \HI observations of
face-on galaxies show that the \HI velocity dispersion is nearly $10
\kms$, even far beyond the optical radius \citep{vanzee99}.  The
timescale for dissolving into the surrounding medium is of the order
of $t_{\rm dis}=R_{\rm S}/\sigma_v$, where $R_{\rm S}$ is the radius
of the shell, and $\sigma_v \approx 10 \kms$ is the velocity
dispersion of HI clouds. For \GSH we find $t_{\rm dis}=18\ \rm
Myr$. \citet{mckee77} also found that the timescale for the
low-density cavity created by a supernova remnant will survive
approximately 10 times longer than the timescale to reach equilibrium
between the internal and external pressure.  Their $t_{\rm max}$ uses
the sound speed in the ambient medium, and is of the same order of
magnitude as $t_{\rm dis}$. The long dissolving timescale resolves the
apparent contradiction of a well-defined \HI shell with expansion
velocity of the order of the velocity dispersion of the ambient
interstellar medium.

\subsection{\GSH compared with SNR models}

Detailed one-dimensional hydrodynamic simulations of the evolution of
supernova remnants including a realistic, metallicity-dependent,
cooling function were made by \citet{thornton98}. A proper treatment
of cooling is essential for simulations that proceed beyond the
adiabatic expansion phase of the remnant. Radiative cooling becomes
important first in the relatively dense expanding shell.  In the late
stages evolution of the supernova shell, the pressure of the hot
interior drives the expansion and compresses the shell further,
increasing the cooling rate in the shell. This leads to a thin neutral
shell in the so-called ``pressure driven snow plow phase''.  This
qualitative scenario is supported by hydrodynamic simulations that
include radiative cooling, e.g. \citet{cioffi88} and
\citet{thornton98}. The pressure driven snow plow phase continues
until the pressure of the hot bubble drops so far that it can no
longer drive the expansion of the remnant.  Not surprisingly, the
timescale for the end of the pressure driven snow-plow phase depends
on the cooling rate of the gas, which itself depends on the density,
metallicity and temperature. \citet{thornton98} included the most
detailed cooling function to date and provide relations for the
properties of the remnant as a function of age, metallicity and
density of the ambient medium.

The density and metallicity of the ambient medium must be assumed in
order to compare the models with the observations.  The metallicity of
the gas is an uncertain factor. Most spiral galaxies show a decline in
metallicity with radius \citep{ferguson98b},
\citep{vanzee98}. Extrapolating the Galactic gradient, with data out
to 18 kpc \citep{smartt97}, to a galactocentric radius of 24 kpc,
results in an oxygen abundance 10\% of the solar value. Similar values
were found for outlying \HII regions in other galaxies (Ferguson
et~al. 1998b). As oxygen lines contribute significantly to the
cooling, the oxygen abundance should be a relevant indicator of the
total metallicity parameter in the models. Therefore, we adopt a
metallicity of 10\% of the solar value.

The scaling relations in \citet{thornton98} were used to obtain the
model predictions listed in Table~\ref{thornton-tab}.  The model
parameters were chosen to cover the relevant range of density and
metallicity and do not represent a fit to the data.  The radius of the
model is particularly sensitive to the density of the ambient medium.
An error of a factor 2 in the density translates into a 50\%
difference in the radius, whereas a factor 10 in metallicity results
in a 20\% difference in radius. The mass of the model shells is
approximately equally sensitive to density and metallicity (20\% for a
factor 2 variation). The models show a very good overall agreement
with the observations if the density of the ambient medium is a factor
$\sim 3$ lower than estimated in Section~\ref{mass-sec} ($n_{\rm
H}=0.27\ \rm cm^{-3}$). The age of the model with $n_{\rm H}=0.09\ \rm
cm^{-3}$ and $\rm log([Z/Z_\sun])=-1$ is approximately 3 Myr. This is
less than the age of \GSH, which is consistent with the smaller
radius, larger expansion velocity and smaller mass of the model.

The main difference between the models and \GSH is in the mass of the
shell, which is a factor 5 larger than the model predictions. However,
in view of the uncertainty in separating emission of \GSH from
unrelated Galactic emission, the distance to \GSH, the energy release
of a supernova explosion adopted in the models, the assumed
metallicity, and the ambient density, this is not a major problem.  We
conclude that the observed parameters of \GSH are consistent with
those of a very old remnant of a single supernova explosion. We note
that occasionally supernovae this far from the Galactic centre must
occur, in view of the evidence for stars as early as spectral type B1
in this part of the Galaxy, e.g. \citet{degeus93}.

The interpretation of \GSH as a supernova remnant makes it the oldest
and the largest supernova remnant known.  This emphasizes the
significance of high-resolution \HI surveys as a means to find the
oldest supernova remnants.  It may not be a coincidence that such an
old but intact supernova remnant is found in the outer galaxy, which
is an environment very different from that near the solar
circle. Individual supernova remnants lose their identity when the
shock speed drops below the sound speed of the interstellar medium, or
when they intersect with neighbouring supernova remnants
\citep{ilovaisky74}.  The timescale for these effects are expected to
be longer in the outer Galaxy because the density of the interstellar
medium is smaller, and supernovae are relatively rare.  Also, the
half-width of the \HI disk at 24 kpc is 1.5 kpc, as opposed to 200 pc
near the solar circle. Therefore, an expanding shell in the outer
Galaxy can grow a factor 5 larger before it bursts out of the disk and
loses its identity as a shell \citep{maclow88,english00}.

Observations of \GSH at wavelengths other than the 21-cm line of \HI
would be of great value to verify its nature. The significance of the
non-detection in the IRAS maps is not clear. Emission by dust in the
foreground is strong, and far infrared emission of the shell may be
weak because of a high gas-to-dust ratio and low interstellar
radiation field in the outer Galaxy. There is no feature at the
position of \GSH in the 34.5 MHz survey of \citet{dwarakanath90}, or
in the 22 MHz survey of \cite{roger99}. However, the upper limit to
the flux density of \GSH at 408 MHz (0.57 Jy) implies a 34.5 MHz flux
density less than 7 Jy at 34.5 MHz, assuming a spectral index of
1. Therefore, the confusion-limited sensitivity of these surveys is
insufficient to expect a detection, and more sensitive observations
are needed. X-ray emission may be expected from a hot bubble inside
the neutral shell. Figure~7 in \citet{thornton98} suggests a
temperature for the interior that is within a factor 3 of $10^6\ \rm
K$. The thermal bremsstrahlung spectrum emitted by the hot interior is
therefore expected to cut off at a photon energy of a few tenths of
keV. With a foregroung HI column density $6.3\times 10^{21}\ \rm
cm^{-2}$, these soft X-rays will be heavily obscured. No optical line
emission is expected because the luminosity of the \citet{thornton98}
models is less than 0.5\% of the peak luminosity in this late stage of
evolution. Arguably the strongest confirmation of a supernova origin
of \GSH would be the identification of a central compact
object. However, no steep-spectrum continuum source could be
identified as a pulsar candidate. A source with an inverted spectrum
was noticed at $(l,b)=(138\fdg198,-1\fdg290)$, but this may well be an
extragalactic object. Finally, the ongoing arcminute resolution \HI
surveys of the Galactic plane provide the best opportunity to identify
similar objects in the far outer Galaxy.

\subsection{Interaction with the environment}

The side of \GSH which faces the Galactic equator is a factor 2
brighter than the opposite side. This asymmetry translates directly into a
factor 2 in density, consistent with a density gradient perpendicular to
the Galactic equator. This raises the question whether the
circular shape of \GSH is consistent with this overall asymmetry. An
estimate of the asymmetry that should be expected can be made by
applying the expansion law in \citet{thornton98}, Equation
(20). Keeping all parameters constant except the ambient density $n$,
we have $R \sim n^{-0.42}$. Assume that the northern part of \GSH
expanded in a medium with a density that is on average twice the
density of the medium on the southern side. Normalizing density and radius
to the values in the direction of the gradient (south), we have $n_1=1$,
$R_1=1$, in the direction of the density gradient.  Scaling
with density, we find $n_2=2$, $R_2=0.747$ in the opposite direction ,
and $n_3=1.5$, $R_3=0.843$ for the direction perpendicular to the
density gradient. We therefore estimate the axial ratio to be $2 R_3 /
(R_1 + R_2)=0.97$. The North-South asymmetry in brightness is
therefore consistent with the nearly circular shape of \GSH. 

The cavities in the upper right quadrant of \GSH discussed in
Section~\ref{substruc-sec} may be smaller, presumably younger, shells.
Cavity 1 is similar in size to the Cygnus Loop. The \HII region found
by \citet{degeus93} is located on the south-eastern edge of Cavity 1.
This is also the region where \GSH appears to make contact with an \HI
cloud (Figure~\ref{chanmap-fig}). The interpretation of Cavity 1 as a
hole in \GSH, and the \HII region on the perimeter of Cavity 1 is
suggestive of progressive star formation. \cite{mckee77} proposed that
the passage of a dense cloud through the shell would leave a hole in
an expanding shell in the snow plow phase, as if casting a shadow.  If
this were the case, we would expect to observe this cloud.  The only
candidate, cloud 2 in \citet{digel94} is much smaller than Cavity 1
(Figure~\ref{C21-fig}). The absence of a radio continuum counterpart
of both Cavity 1 and Cavity 2 argues against the interpretation that
these are young supernova remnants. It was argued in Section~\ref{source-sec}
that the non-detection of radio continuum emission does not rule out
that it is a supernova remnant because of its large age.

\subsection{Implications for galaxies}

We have argued that the outer Galaxy is a less hazardous environment
for supernova remnants because radiative cooling proceeds at a lower
rate, the supernova rate is lower, and the scale-height of the gas
disk is larger, than in the inner Galaxy. A similar environment is
usually found in the outskirts of other spiral galaxies, and almost
everywhere in dwarf irregular and low surface brightness galaxies.
\GSH may be the prototype of a class of supernova remnant that is
representative for such a low metallicity, low density environment.
The timescale to dissolve completely into the interstellar medium when
the expansion halts, $t_{\rm dis}=R_{\rm S}/\sigma_v$ is of the order of
10 Myr.  Assuming this process recently started in \GSH, at the age of
4.3 Myr, it becomes clear that the shell will remain as large density
enhancement between 5 and 10 Myr after the supernova explosion. We
speculate that such old supernova shells may be an important source of
dense clouds in a low-density environment. This allows star formation
to progress for a long time after a single trigger event.

An expanding shell like \GSH would not be resolved in most studies of
the \HI velocity dispersion in galaxies outside the local group, but
it would fill a significant part of the beam. This may lead to a
broader local \HI profile, without a counterpart in the optical or
radio continuum.

\section{Conclusions}

\GSH is a large \HI shell at a Galactocentric radius of 24 kpc
(heliocentric distance 16.6 kpc). Its radius ($180 \pm 10\ \rm pc$)
and expansion velocity ($11.8 \pm 0.9 \kms$) suggest an age of 4.3
Myr, assuming the expansion law $R\sim t^{2/7}$. The total mass of the
shell is $2 \times 10^{5}\ \rm M_\sun$, and the kinetic energy of the
expansion is $3 \times 10^{43}\ \rm J$. Stellar wind and a supernova
explosion are considered as possible sources of the shell.

If \GSH is a stellar wind bubble, the expansion energy of \GSH sets a
strong lower limit $60\ \rm M_\sun$ to the mass of the star if the
expansion is driven only by a stellar wind. The \HII region associated
with such a star would have been detected in the radio continuum map
at 1420 MHz, or optically.  There is also no evidence for a young
supernova remnant inside the \HI shell.

These objections do not exist if \GSH is a supernova remnant. The
properties of \GSH correspond well with hydrodynamic models of
supernova remnants in a late stage of evolution. \GSH is then the
oldest and largest supernova remnant known. Its large size and age may
be related to a low-density, low-metallicity environment.

\acknowledgments{This work has been supported by the Natural Sciences
and Engineering Research Council of Canada (NSERC) for JI, and also by
NSERC support towards Canadian Galactic Plane Survey postdoctoral
fellowships. JMS is guest user, Canadian Astronomy Data Centre, which
is operated by the Herzberg Institute of Astrophysics, National
Research Council of Canada. The authors thank the anonymous referee for
useful comments on the manuscript.}


\begin{thebibliography}{}

\bibitem[Abbot(1982)]{abbot82} Abbot, D. C. 1982, \apj, 263, 723
\bibitem[Bransford et~al.(1999)]{bransford99} Bransford, M. A., Thilker, D. A., Walterbos, R. A. M., \& King, N.L. 1999, \aj, 118, 1635
\bibitem[Blair et~al.(1981)]{blair81} Blair, W. P., Kirshner, R. P., \& Chevalier, R. A. 1986, \apj, 247, 879
\bibitem[Bressan et~al.(1993)]{bressan93} Bressan, A., Fagotto, F., Bertelli, G., \& Chiosi C. 1993, A\&AS, 100, 647
\bibitem[Brown et~al.(1995)]{brown95} Brown, A. G. A., Hartmann, \& D., Burton, W. B. 1995, \aap, 300, 903
\bibitem[Burton(1991)]{burton91} Burton, W. B. 1991, in The Galactic Interstellar medium, Saas-Fee course 21, eds. Burton W.B., Elmegreen B.G., Genzel R.
\bibitem[Carpenter(1997)]{carpenter97} Carpenter, J. M. 1997, \apjs, 115, 241
\bibitem[Cioffi et~al.(1988)]{cioffi88} Cioffi, D. F., McKee, C. F., \& Bertschinger, E. 1988, \apj, 334, 252
\bibitem[Digel et~al.(1994)]{digel94} Digel, S., De Geus, E., \& Thaddeus, P. 1994, \apj, 422, 92\
\bibitem[Dwarakanath \& Undaya Shankar(1990)]{dwarakanath90} Dwarakanath, K. S., Undaya Shankar, N. 1990, JApA, 11, 323  
\bibitem[English et~al.(2000)]{english00} English, J., Taylor, A. R., Mashchenko, S. Y., Irwin, J. A., Basu, S., \& Johnstone, D. 2000, \apjl, 533, L25
\bibitem[Ferguson et~al.(1998a)]{ferguson98a} Ferguson, A. M. N., Wyse, R. F. G., Gallagher, J. S., \& Hunter, D. A. 1998a, \apjl, 506, L22
\bibitem[Ferguson et~al.(1998b)]{ferguson98b} Ferguson, A. M. N., Gallagher, J. S., \& Wyse, R. F. G. 1998b, \aj, 116, 673
\bibitem[De Geus et~al.(1993)]{degeus93} De Geus, E. J., Vogel, S., Digel, S. W., \& Gruendl, R. A. 1993, \apjl, 413, L97
\bibitem[Gordon et~al.(1999)]{gordon99} Gordon, S. M., Duric, N., Kirshner, R. P., Goss, W., \& Viallefond, F. 1999, \apjs, 120, 247
\bibitem[Green(1984)]{green84} Green, D. A. 1984, \mnras, 209, 449
\bibitem[Green(1988)]{green88} Green, D. A. 1988, \apss, 148, 3
\bibitem[Green(1991)]{green91} Green, D. A. 1991, \pasp, 103, 209
\bibitem[Heiles(1984)]{heiles84} Heiles, C. 1984, \apjs, 55, 585
\bibitem[Heyer et~al.(1998)]{heyer98} Heyer, M. H., Brunt, C., Snell, R. L., Howe, J. E., Schloerb, F. P., \& Carpenter, J. M. 1998, \apjs, 115, 241
\bibitem[Higgs et~al.(1999)]{higgs99} Higgs, L. A. 1999, in ASP Conf. Ser. 168, New Perspectives on the Interstellar Medium, ed. A. R. Taylor, T. L. Landecker, G. Joncas (San Francisco: ASP), 15
\bibitem[Hunter et~al.(1998)]{hunter98} Hunter, D. A., Wilcots, E. M., Van Woerden, H., Gallagher, J. S., \& Kohle, S. 1998, \apjl, 495, L47
\bibitem[Hunter et~al.(1999)]{hunter99} Hunter, D. A., Van Woerden, H., \& Gallagher, J. S. 1999, \aj, 118, 2184
\bibitem[Ilovaisky et~al.(1974)]{ilovaisky74} Ilovaisky, S. A., \& Lequeux, J. 1974, \aap, 20, 347
\bibitem[Kuijken \& Tremaine(1994)]{kuijken94} Kuijken, K., \& Tremaine, S. 1994, \apj, 421, 178
\bibitem[Lamers et~al.(1999)]{lamers99} Lamers, J. G. L. M., Haser, S., De Koter, A., \& Leitherer, C. 1999, \apj, 516, 872  
\bibitem[Landecker et~al.(1990)]{landecker90} Landecker, T. L., Clutton-Brock, M., \& Purton, C. R. 1990, \aap, 232, 207
\bibitem[Lanning(1977)]{lanning77} Lanning, H. H. 1973, \pasp, 85, 70
\bibitem[Leitherer et~al.(1992)]{leitherer92} Leitherer, C., Robert, C., \& Drissen, L. 1992, \apj, 401,596 
\bibitem[Levenson et~al.(1998)]{levenson98} Levenson, N. A., Graham, J. R., Keller, L. D., \& Richter, M. J. 1998, \apjs, 118, 541
\bibitem[Matthews(1998)]{matthews98} Matthews, B. C., Wallace, B. J., \& Taylor, A. R. 1997, \apj, 493, 312
\bibitem[Merrifield(1992)]{merrifield92} Merrifield, M. R. 1992, \aj, 103, 1552
\bibitem[Mac Low \& McCray(1988)]{maclow88} Mac Low, M.-M., \& McCray, R. 1988, \apj, 324, 776
\bibitem[McKee \& Ostriker(1977)]{mckee77} McKee, C. F., \& Ostriker, J. P. 1977, \apj, 218, 148
\bibitem[Normandeau(1999)]{normandeau99} Normandeau M., 1999, AJ 117,2440
\bibitem[Normandeau et~al.(2000)]{normandeau00} Normandeau M., Taylor, A.R., Dewdney, P. E., \& Basu, S. 2000, AJ 119,2982
\bibitem[Nugis \& Lamers(2000)]{nugis00} Nugis, T., \& Lamers, H. J. G. L. M. 2000, \aap, 360, 227
\bibitem[Pineault \& Joncas(2000)]{pineault00} Pineault, S., \& Joncas, G. 2000, \aj, 120,3218
\bibitem[Routledge et~al.(1986)]{routledge86} Routledge, D., Landecker T. L., \& Vaneldik, J. F. 1986, \mnras, 221, 809
\bibitem[Roberts(1972)]{roberts72} Roberts, W. W. 1972, \apj, 173, 259
\bibitem[Roger et~al.(1999)]{roger99} Roger, R. S., Costain, C. H., Landecker, T. L., \& Swerdlyk, C. M. 1999, A\&AS, 137, 7 
\bibitem[Sellwood \& Balbus(1999)]{sellwood99} Sellwood, J. A., \& Balbus, S. A. 1999, \apj, 511, 660 
\bibitem[Smartt \& Rollestone(1997)]{smartt97} Smartt, S. J., \& Rolleston, W. R. J. 1997, \apjl, 481, L47
\bibitem[Taylor et~al.(1999)]{taylor99} Taylor, A. R. 1999, in ASP Conf. Ser. 168, New Perspectives on the Interstellar Medium, ed. A. R. Taylor, T. L. Landecker, G. Joncas (San Francisco: ASP), 3
\bibitem[Thornton et~al.(1998)]{thornton98} Thornton, K., Gaudlitz, M., Janka, H.-Th., \& Steinmetz, M. 1998, \apj, 500, 95
\bibitem[Vink et~al.(2001)]{vink01} Vink, J. S., De Koter, A., \& Lamers, H. J. G. L. M. 2001, \aap, 369, 574
\bibitem[Van Zee et~al.(1998)]{vanzee98} Van Zee, L., Salzer, J. J., Haynes, M. P., O'Donoghue, A. A., \& Balonek, T. J. 1998, \aj, 116, 2805
\bibitem[Van Zee \& Bryant(1999)]{vanzee99} Van Zee, L., \& Bryant, J. 1999, \aj, 118, 2172
\end{thebibliography}
\end{document}